\documentclass{article}
\usepackage{ijcai24}

\usepackage{times}
\usepackage{soul}
\usepackage{url}
\usepackage[hidelinks]{hyperref}
\usepackage[utf8]{inputenc}
\usepackage[small]{caption}
\usepackage{graphicx}
\usepackage{amsmath}
\usepackage{amsthm}
\usepackage{booktabs}
\usepackage{algorithm}
\usepackage{algorithmic}
\usepackage[switch]{lineno}

\usepackage{xspace}
\usepackage{amsmath,amssymb}
\usepackage{bm}
\usepackage{booktabs}
\usepackage{multirow}
\usepackage{microtype}

\usepackage{newfloat}
\usepackage{listings}

\title{Perception-Inspired Graph Convolution for Music Understanding Tasks}

\author{
Emmanouil Karystinaios$^1$
\and
Francesco Foscarin$^{1,2}$
\and
Gerhard Widmer$^{1,2}$
\affiliations
$^1$Institute of Computational Perception, Johannes Kepler University Linz, Austria\\
$^2$LIT AI Lab, Linz Institute of Technology, Austria\\
\emails
\{emmanouil.karystinaios, francesco.foscarin, gerhard.widmer\}@jku.at
}

\usepackage{xcolor}

\begin{document}

\newcommand{\Rel}{\mathcal{R}}
\newcommand{\V}{V}
\newcommand{\Ein}{\ensuremath{E}}
\newcommand{\Epred}{\ensuremath{E_{\textrm{pred}}}}
\newcommand{\Etarg}{\ensuremath{E_{\textrm{target}}}}
\newcommand{\round}[1]{\ensuremath{\lfloor#1\rceil}}

\newcommand{\notimplies}{\;\not\!\!\!\implies}

\newcommand{\PotE}{\ensuremath{\Lambda}}
\newcommand{\our}{MusGConv\xspace}
\newcommand{\ouref}{MusGConv(+EF)\xspace}
\newcommand{\RelOns}{\ensuremath{e^\textrm{onset}_{vu}}}
\newcommand{\RelDur}{\ensuremath{e^\textrm{dur}_{vu}}}
\newcommand{\RelPitch}{\ensuremath{e^\textrm{pitch}_{vu}}}
\newcommand{\PCInt}{\ensuremath{\bm{e}_{vu}^{\textrm{PCInt}}}}

\newcommand{\relF}{Distance \xspace}
\newcommand{\PCIntF}{PCInt \xspace}

\maketitle

\begin{abstract}


We propose a new graph convolutional block, called \our, specifically designed for the efficient processing of musical score data and motivated by general perceptual principles. It focuses on two fundamental dimensions of music, pitch and rhythm, and considers both relative and absolute representations of these components. We evaluate our approach on four different musical understanding problems: monophonic voice separation, harmonic analysis, cadence detection, and composer identification which, in abstract terms, translate to different graph learning problems, namely, node classification, link prediction, and graph classification. Our experiments demonstrate that \our improves the performance on three of the aforementioned tasks while being conceptually very simple and efficient.
We interpret this as evidence that it is beneficial to include perception-informed processing of fundamental musical concepts when developing graph network applications on musical score data.
All code and models are released on \url{https://github.com/manoskary/musgconv}.
\end{abstract}


\section{Introduction}\label{sec:intro}


Music data can be represented in computer applications in multiple formats, two popular ones being audio and symbolic representations. The first encodes a measure of the pressure/intensity of the sound wave over time,  while the second explicitly encodes discrete musical events such as notes and rests (see \cite{foscarin2022match} for an overview of different symbolic music formats). Due to this higher-level information, symbolic representations are generally considered better inputs/outputs for music analysis and generation tasks in the Music Information Research (MIR) field. Moreover, any musical task that starts from a musical score (sheet music) or from MIDI files is naturally in the symbolic domain. 

In the MIR literature, symbolically encoded music is typically handled with techniques heavily inspired by computer vision (CV) or natural language processing (NLP) research. 
In the first case, music is represented in a so-called "piano roll" format (first developed in MIDI sequencers) and treated as a raster image, with the X axis being time, and the Y axis pitch.
The pitch values are typically encoded as MIDI pitch, i.e., with integers in $[0-127]$, which covers all notes that can be played by common well-tempered instruments (a range that is larger than the range of the piano, with 88 notes), and the time resolution is a parameter that is usually set to the expected shortest note duration. In its simpler version, each element of the 2D matrix is set to 1 if a note is sounding at the corresponding pitch and time, or 0 otherwise. The downside of this approach is that it creates a very large and very sparse input matrix since only a few notes will play at any time.
The other common approach is treating music with sequential models from NLP research. Although different tokenization techniques have been proposed~\cite{miditok2021}, it is easy to argue that music does not fit well into strictly sequential models, since more than one note can sound simultaneously, notes can partially overlap, and generally, the pitch-temporal relations between notes hold important musical information.

A few recent works~\cite{hsiao2021learning,jeong2019graph,karystinaios2022cadence,karystinaios2023voice,huan} started to explore the use of graphs, and Graph Neural Networks (GNNs), to represent and process symbolic music. Significant advances have been reported on various musical tasks~\cite{karystinaios2023voice,karystinaios2023roman}. However, the components that these papers use are ``borrowed'' from GNN research with other types of data. 

We argue that this can lead to suboptimal results. As a solution, we design \our, a new graph convolutional block specifically dedicated to music data, that is founded on fundamental music perception principles.
We test our approach on four musical tasks: voice separation, composer classification, Roman numeral analysis, and cadence detection. This selection of tasks allows us to cover three major graph neural network classes of problems: graph classification, link prediction, and node classification.
We compare our results with the state-of-the-art graph models for these tasks and show that the use of \our leads to better results overall. Moreover, its simple design enables this performance boost without adding any additional computational cost.
%


\section{Perceptual and Modeling Considerations}\label{sec:perceptual}

We base our research on the perceptual principles of two fundamental musical dimensions: pitch and rhythm.
Cognitive studies show that people are not very sensitive to the absolute pitch of individual notes and perceive mainly the distance between pitches \cite{deutsch2013psychology}. Thus, we perceive the same musical pattern if it is shifted higher or lower in frequency. This is called \textit{relative pitch perception} and has been formalized in music theory through measures of pitch distance called \textit{intervals}, which are the basis of all musical concepts that involve combinations of pitches, such as chords and harmony.
The position (and duration) of notes in time is also not important in itself but only relative to the position (and duration) of the other notes. This temporal organization is called \textit{rhythm} and is typically composed of patterns that tend to be periodic and organized at different hierarchical levels. 

\begin{figure}[t]
\centerline{\includegraphics[width=\columnwidth]{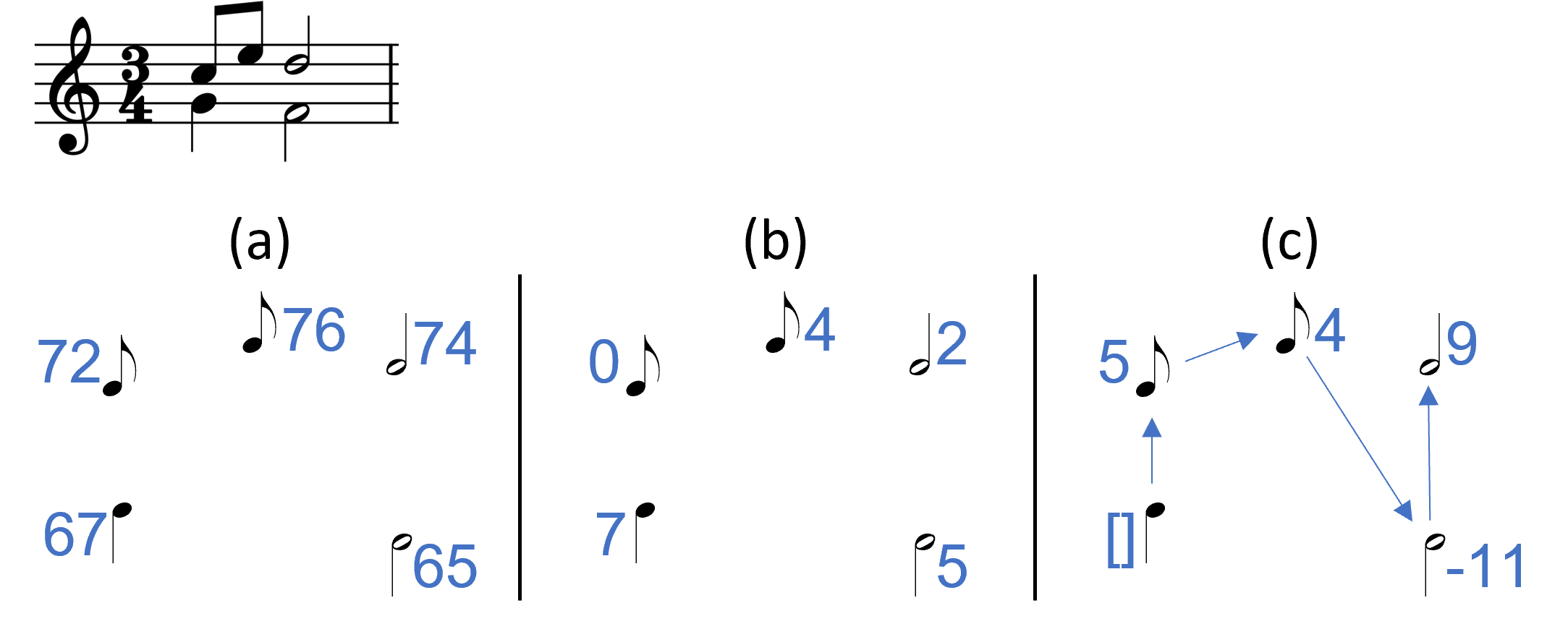}}
\caption{Three alternative representations of note pitches in a musical excerpt: (a) absolute representation in terms of MIDI pitch; (b) relative pitch distance (ignoring the octave) in semitones relative to the fundamental pitch specified by the key signature (here: C); (c) relative pitch distance in semitones from the closest preceding note; in case of chords the order is defined from bottom to top.}
\label{fig:abs_rel_repr}
\end{figure}

While these principles are simple, producing relative features to use as input for deep learning systems is not an easy task.
For the pitch representation, considering intervals between consecutive notes instead of absolute pitches, as proposed for example in the IDyOM framework \cite{Pearce:2018,Pearce:idyom}, poses the problem of defining a note order. This is trivial for monophonic melodies but becomes problematic for polyphonic music, where multiple notes can fully or partially overlap in time. Figure~\ref{fig:abs_rel_repr} (c) exemplifies a possible ordering rule, but such rules may not be generally valid for different pieces and contexts
or in general yield very different interval sequences for very similar patterns (e.g., if we remove the F), thus not helping with learning a general representation. Inserting interval information after the music has been tokenised in a sequential representation \cite{kermarec2022improving}, creates similar ordering problems.

Another strategy is to rely on music theory and consider pitch distances relative to the fundamental note defined by the key of the piece (see Fig.~\ref{fig:abs_rel_repr} (b). This has a musicological limitation, as the key signature of a piece may not change when there are temporary modulations to other keys, and a practical one, as keys are not always notated in musical datasets and MIDI files. Not to mention all the music that falls outside the classic tonal framework, for which the definition of a key is not even meaningful.
The most common alternative, e.g., \cite{nakamura2016tree,foscarin2021pkspell}, has been the use of data augmentation via transposition, assuming that if the network sees transpositions of the same piece, it will learn patterns that are general across all transpositions. However, this is far from ideal since the network will need to store similar patterns redundantly, thus making inefficient use of network capacity and drastically increasing training time. 


For this reason, we believe that designing relative features to input to our system is not a viable option for general music modelling. 
We explore a different path, which is to customize the working mechanism of our network to take into account the relative perceptual properties of music through a \textit{dedicated message-passing mechanism} that computes pairwise pitch and time representations.
This
is weakly related to
recent work on audio representations of music that aims at learning transposition-invariant features~\cite{Lattner2018LearningTI,Elowsson2019ModelingMM,Lattner2019LearningCB} and tempo-invariant rhythmic patterns~\cite{Giorgi2021DownbeatTW}. Other related work targets graphs whose edges encode geometric information \cite{Atzmon2018,WangSLSBS19,Satorras2021EnEG}, intending to build representations that are invariant to operations such as translation. The ideas from these last approaches (with minor modifications) are also beneficial for musical tasks, but we show that our music-specific approach outperforms them.

Up to this point, we have highlighted the importance of relative pitch and time representations. However, their absolute values could also be important depending on the task at hand. For example, instruments would peculiarly change their timbre as they approach very low or very high notes in their range, which is recognisable by a listener. In tonal music, the absolute pitch of notes defines a key signature which could be relevant in the composer classification task (specific keys can have very specific meanings to composers, and within a musical tradition). The same goes for absolute time positions, for example, with patterns happening at the beginning or end of a piece. Finally, the output of our network may need to be an absolute pitch (for example, in the Roman numeral analysis task we describe later 
), and therefore we need to retain this information in the network.

This need for considering both absolute and relative pitch and time positions motivates the design of our new convolutional block, to be described in detail in Section~\ref{sec:method} below.

\section{Graph Approaches to Musical Tasks}\label{sec:tasks}

\begin{figure*}[tbp]
\centerline{\includegraphics[width=\textwidth]{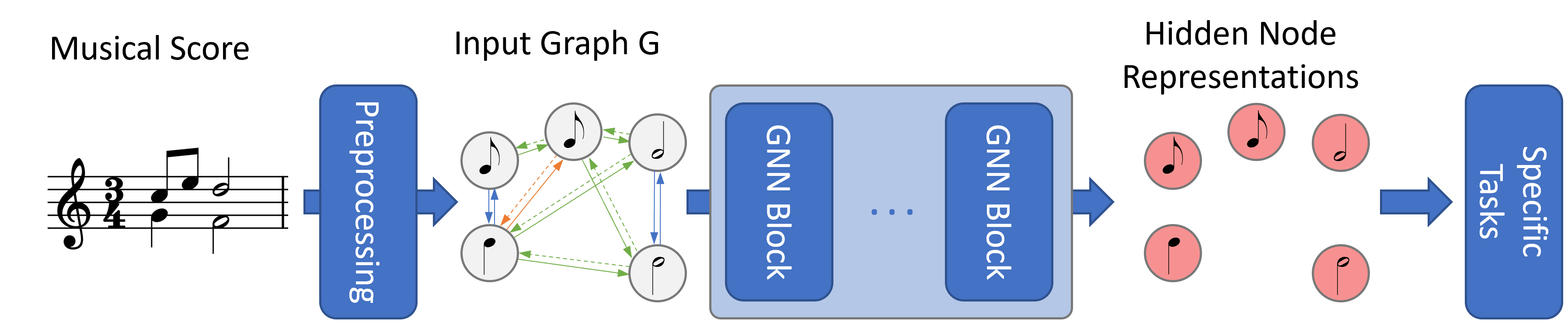}}
\caption{General architecture of our pipeline. The first part that produces the hidden node representation is common among all tasks; the last module is task-specific.}
\label{fig:general}
\end{figure*}

In this section, we describe existing graph modelling approaches to the four musical tasks we use to evaluate our proposal. 
They all have a common pipeline which involves building a graph from a given musical score (see Figure~\ref{fig:general}) and using a series of convolutional blocks to produce context-aware hidden representations for each node. We start by describing the graph-building procedure and a generic graph convolutional block; we then proceed by detailing the tasks and the specific network components used to target them.

\subsection{Graph from Musical Scores}\label{subsec:previous}

A graph is defined as a set of nodes $V$ and a set of edges $E$, where each edge $(u,v) \in E$ connects the nodes $v,u \in V$.
We extend this definition by considering labelled edges which we model as a triple $(u, r, v)$, where $u, v \in V$ and $r \in \mathcal{R}$ is a relation type. A graph with edges of multiple types (we don't use different types of nodes in this work) is called \textit{heterogeneous}.
Moreover, we consider an \textit{attributed} graph, i.e., a graph where each node is described by a set of features, which we group in the columns of the matrix $X$.
Our attributed heterogeneous graph is defined as $G = (V, E, X)$.

We create such a directed graph from a musical score following the work of Karystinaios et al.~\shortcite{karystinaios2023voice}.
Each node $v \in V$ corresponds to one and only one note in the musical score. $\mathcal{R}$ includes 4 types of relations: onset, during, follow, and silence, corresponding, respectively, to two notes starting at the same time, a note starting while the other is sounding, a note starting when the other ends, and a note starting after a time when no note is sounding. The inverse edges for during, follows, and silence relations are also created.

The feature matrix $X$ is composed of the following features extracted from each note of the score: the pitch class, i.e., one of the 12 note names (C, C\#, D, D\#, etc.), the octave in $[1,\dots,7]$, the note duration, encoded as a single float value $d \in [0,1]$ computed as the ratio of the note and bar durations, passed through a tanh function to limit its value and give more resolution to shorter notes, as proposed by Karystinaios et al.~\shortcite{karystinaios2023voice}.
For the task of Roman Numeral Analysis and Cadence Detection, we add additional specialized features to be consistent with the approach in the literature we consider in our evaluation~\cite{karystinaios2023roman,karystinaios2022cadence}.

\subsection{Graph Convolution Operation}\label{sec:lit_graph_conv}

We now present a generic graph convolutional block, to simplify the description of our music-dedicated approach in the next section.
Given an attributed homogeneous graph $G$, a graph convolution block  that updates the representation of node $u$ for layer $l+1$ can be described as:
\begin{align}
    \bm{h}^{(l+1)}_u &= \psi \left( \bm{h}^{(l)}_u,
\underset{v \in \mathcal{N}(u)}{\textrm{aggregate}} \, (\{\bm{\eta}_{vu}^{(l)}\}) \right), \label{eq:psi}\\
\bm{\eta}_{vu}^{(l)} &= \phi \left(\bm{h}^{(l)}_v, \bm{h}^{(l)}_u \right) \label{eq:eta}\\
    \bm{h}^{(l+1)}_u &= \sigma \left( \bm{h}^{(l+1)}_u  \right)\label{eq:nonlin}
\end{align}
where $\textrm{aggregate}(\cdot)$ denotes a differentiable, permutation invariant aggregation function, e.g., sum, mean, etc.; $\phi$ and $\psi$ are called edge operation and node operation, respectively, 
and denote differentiable learnable functions such as concatenation, sum, or multiplication, followed by a linear transformation; 
$\sigma$ denotes a non-linear function, $\mathcal{N}(u)$ denotes the neighbours of $u$; $\bm{h}_u^{(l)}$ is the hidden representation of node $u$ at layer $l$.

Furthermore, if we want to leverage edge features, Equation~\ref{eq:eta} becomes:

\begin{equation}\label{eq:eta_edge_features}
    \bm{\eta}_{vu}^{(l)} = \phi \left(\bm{h}^{(l)}_v, \bm{h}^{(l)}_u, \bm{e}_{vu}\right)
\end{equation}
\noindent
where, $\bm{e}_{uv}$ are features of the directed edge connecting node $v$ with node $u$.

When $G$ is heterogeneous, the function $\mathcal{N}(u)$ in Equation~\ref{eq:psi} is modified as proposed in \cite{SchlichtkrullKB18} to return only the neighbours nodes which are connected with an edge of type $r$. Equations~\ref{eq:psi} and \ref{eq:nonlin} are then computed $|\mathcal{R}|$ times and the results $\mathbf{h_r}^{(l)}_{u}$ aggregated in an unique node latent representation $\bm{h}^{(l)}_u$ as:

\begin{equation}\label{eq:aggregation_heterogeneous}
    \bm{h}^{(l)}_u = \underset{r 
    \in \Rel}{\textrm{aggregate}}\left( \{ \mathbf{h_r}^{(l)}_{u} \} \right)
\end{equation}
\noindent
where $\textrm{aggregate}(\cdot)$  denotes a differentiable, permutation invariant aggregation function such as sum or mean. 

We now turn to describe the tasks that we will use for evaluation and the task-specific part of our graph model.

\subsection{Monophonic Voice Separation}\label{subsec:vocsep}

Voice separation is the task of segmenting a symbolic music piece into an unknown number of individual monophonic note streams according to musical and perceptual criteria.
Duoane and Pardo~\shortcite{duane2009streaming} framed the problem as a link prediction task in which two notes are linked if they are consecutive in the same voice. Karystinaios et al.~\shortcite{karystinaios2023voice} proposed a GNN-based model that reached new state-of-the-art results.

Following their approach, we perform this task by adding a link predictor module, consisting of an MLP which takes as input the hidden representations of nodes and for each pair of nodes performs a binary classification between the ``linked'' and ``not-linked'' classes.

The evaluation metric is the \textit{binary F1 score}, i.e. the F1 score for the positive class which represents the true links in the ground truth.
Karystinaios et al. also consider the F1 score after a postprocessing phase, but we don't use it to keep the number of metrics of reasonable size, and because, as they report, postprocessing increases the metric in a way which does not always correlate perfectly with the network performance.

\subsection{Composer Classification}\label{subsec:composer}
Composer classification from a symbolic musical score is the task of identifying the composer of the score from a list of composers. In the graph problem taxonomy, this falls in the category of global graph classification tasks. 
Following the work of Zhang et al.~\shortcite{huan}, we perform this task by adding a global mean pooling layer to our architecture, which averages the latent representations built by our GNN blocks, followed by an MLP which predicts probabilities over composer classes.
The composer classification predictions are evaluated in terms of \textit{classification accuracy}.

\subsection{Roman Numeral Analysis}\label{subsec:rna}
Roman numeral analysis is a branch of analytical musicology whose goal is to infer the underlying harmony and chord progressions from a musical score. The result is a set of complex composite labels (the Roman numerals) which annotate music at the onset level, i.e., for every score position that corresponds to one or more note onsets.
Related work~\cite{chen2018functional,micchi2020not,mcleod2021modular,lopez2021augmentednet} frames RN analysis as a multi-task classification problem where every label is broken down into 5 components (degree, inversion, root, key, and quality) which are predicted  by different classifiers in 
hard-parameter-sharing setting. 
Of these components, degree, inversion, and quality are transposition invariant and the rest depend on the absolute pitches in the input. 
A recent approach~\cite{karystinaios2023roman} considers a graph input and obtains new state-of-the-art results.

Following this work, we perform this task by adding to our general architecture an onset edge pooling layer that contracts the latent representations from the note-wise level to the onset-wise, i.e., it creates a single vector per unique onset. All vectors are then ordered by time position and fed into a GRU layer whose output is finally used by the aforementioned MLP classifiers.
In the taxonomy of graph problems, RN analysis falls in between node classification and subgraph classification because of the effect of the edge pooling layer. 

The evaluation score is the so-called \textit{Chord Symbol Recall (CSR)}, i.e., the ratio of the total duration of segments where prediction equals annotation vs.~the total duration of annotated segments~\cite{harte2010towards}. 

\subsection{Cadence Detection}\label{subsec:cadence}
Cadence detection is a music analysis task that consists of detecting cadences, i.e. phrase endings with a strong and specific melodic-harmonic closure effect, in a musical score. 
Cadences are important both musicologically and perceptually and it is known that they relate to particular voicings and chord progressions; however, their automatic predictions remain particularly challenging due to the high number of exceptions and corner cases.
A recent graph approach \cite{karystinaios2022cadence} framed the problem as a multiclass node classification scenario, by predicting the presence of a cadence and its type for each note. 

Following that work, we use a graph autoencoder architecture and the latent synthetic oversampling technique SMOTE~\cite{chawla2002smote} to balance the heavily unbalanced class labels.
For the same reason, the reported evaluation metric is the \textit{macro F1 score}.

\section{Our Approach: \our}\label{sec:method}

\begin{figure}[tbp]
\centerline{\includegraphics[width=0.7\columnwidth]{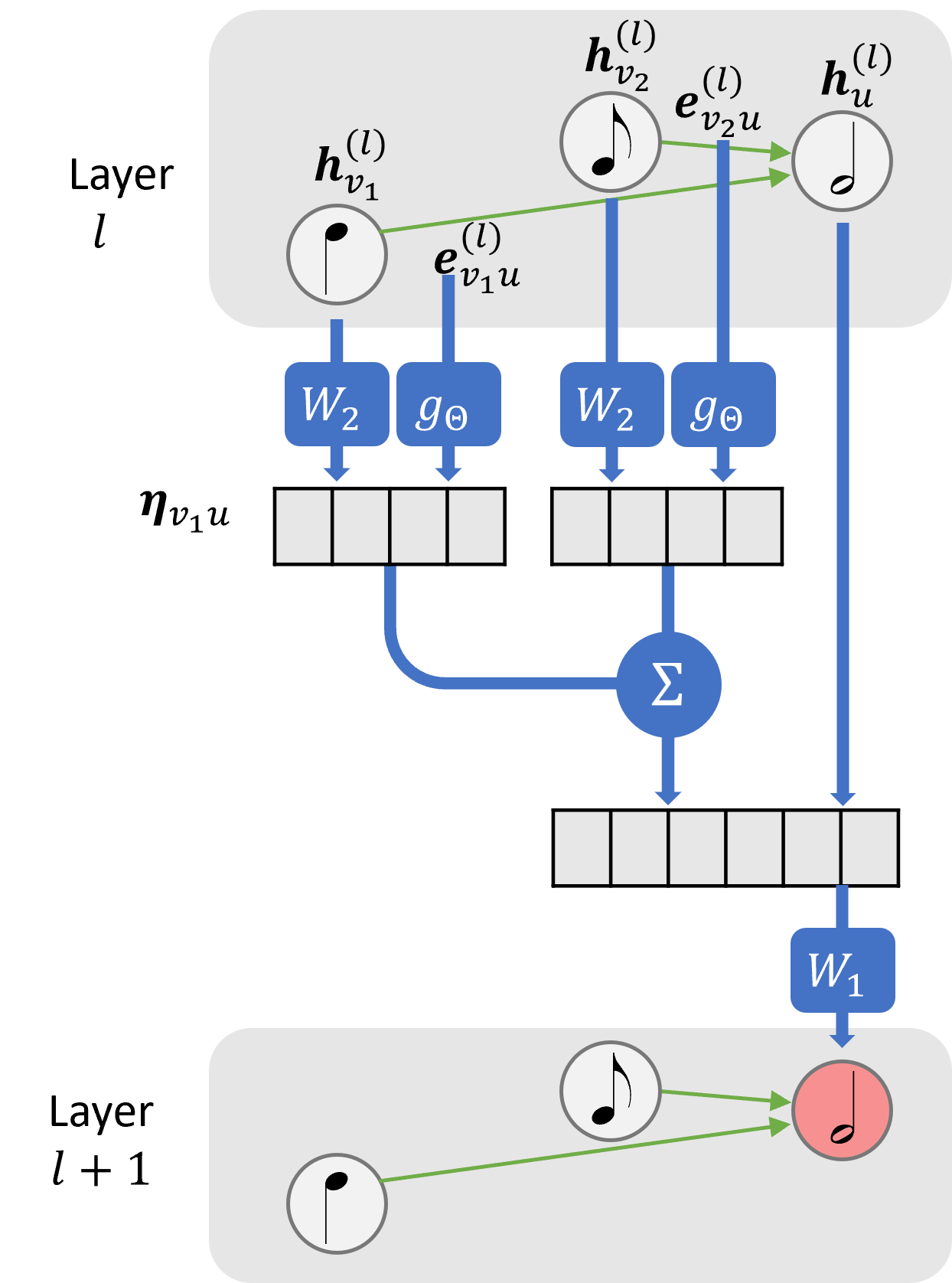}}
\caption{Visualization of update for node $u$ in our \our block (considering only one edge type), corresponding to Eqns.~\ref{eq:our_mespas} and \ref{eq:our_eta}.}
\label{fig:musicconv}
\end{figure}

Similarly to previous works, we use stacked graph convolutions to create a hidden representation of notes that is then used as input for specific music tasks (see Figure~\ref{fig:general}).
In our proposed approach, we replace the graph convolutional blocks in the stack with our novel convolutional block (see Figure~\ref{fig:musicconv}). 

In terms of the notation introduced in the general description of the graph convolution in Section~\ref{sec:lit_graph_conv}, our convolutional block is characterised by two core contributions: a way to build the edge features $\bm{e}$, and the choice of the edge operation $\phi$ in Equation~\ref{eq:eta_edge_features}. 

\subsection{Edge Features Computation}\label{sec:edge_feat_comp}
For each edge between nodes $u,v$, we consider three edge features: $\RelOns, \RelDur, \RelPitch$ each of them encoded as a single scalar corresponding to the distance between onset, duration, and pitch, respectively.
\begin{align}\label{eq:relational_features}
    \RelOns &= | on(u) - on(v) | \nonumber \\
    \RelDur &= | dur(u) - dur(v) | \\
    \RelPitch &= | pitch(u) - pitch(v)| \nonumber
\end{align}

This roughly corresponds to the computation of distance in papers that deal with geometrical data~\cite{Satorras2021EnEG} except that in their case it is a multidimensional space distance, while we compute a set of one-dimensional distances since it makes no sense to mix duration, pitch and onset information.
To keep these values in a convenient numerical range, we normalise each feature with $\ell_2$ normalisation over all edges in a batch. 

Additionally, we inform the network about the \textit{pitch-class interval} (PCInt), i.e., the distance between notes without considering the octave, or the interval direction. This can be seen as a relative version of the \textit{chroma feature}, which is commonly used in MIR tasks related to the harmonic content of the music.
This integer in $[0,\dots,11]$ is passed through an \textit{embedding layer}, i.e., a learnable look-up table which maps these integers to points $\PCInt$ in a continuous multidimensional space.

For each edge, all the aforementioned edge features are concatenated in a single vector:
\begin{equation}\label{eq:fullfeat}
    \bm{e}_{uv}^{(0)} = \textrm{cat}\left(\RelOns,\RelDur,\RelPitch,\PCInt \right)
\end{equation}

\subsection{Edge Operation}\label{subsec:edge_operation}

Our second contribution, needed to properly leverage the edge feature information, is a new formulation of the message passing paradigm (see Figure~\ref{fig:musicconv} for a graphical representation).
The new edge operation defines the edge operation $\phi$ in Equation~\ref{eq:eta_edge_features} as follows:
\begin{equation}
\bm{\eta}^{(l)}_{vu} = \textrm{cat}\left(\bm{W}^{(l)}_2 \bm{h}^{(l)}_v, g_\mathbf{\Theta}^{(l)} \left(\bm{e}^{(l)}_{vu}\right)\right) \label{eq:our_eta}
\end{equation}
\noindent
where $g_\mathbf{\Theta}$ denotes a simple two-layer MLP with a Relu activation and layer-wise normalization.

\begin{figure}[tbp]
\centerline{\includegraphics[width=0.7\columnwidth]{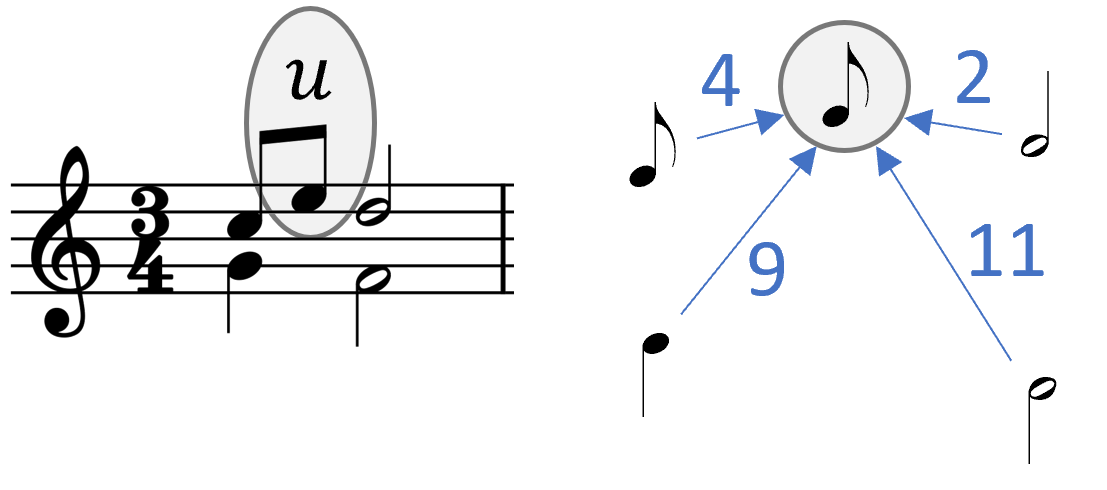}}
\caption{Relative Pitch features $\RelPitch$ for the highlighted note $u$.}
\label{fig:rel_repr}
\end{figure}

Note how we concatenate the edge features here, while other approaches that deal with edge features tend to apply a permutation-invariant operation such as multiplication or addition~\cite{Satorras2021EnEG,hu2019strategies}. In this way, the transposition- and time-invariant information carried by the edge features is treated in the same way as the node feature when computing the pairwise representation $\eta_{uv}$. This is inspired by musicological considerations: we don't want to weight/modify the absolute representation according to the relative representation, but rather to just use it as input, as is done by cognitively plausible musical models~\cite{Pearce:2018,Pearce:idyom}. This is similar to what we discuss in Section~\ref{sec:intro} but, by using \our, we no longer have the problem of setting a (debatable) order of the input notes, since for each node, we will consider the relative features according to every other node connected to it (see Figure~\ref{fig:rel_repr}).

We consider two variants of our system which differ in the edge features which are passed to layers after the first. The first variant, named \our uses the absolute difference of the node hidden embeddings, i.e. $\forall l>0$,
\begin{equation}\label{eq:edgesub}
 \bm{e}^{(l)}_{vu} = |\bm{h}_v^{(l)} - \bm{h}_u^{(l)}|
\end{equation}
The second variant, named \ouref, uses the edge hidden representation from the previous layer as the edge features of the next layer: 
\begin{equation}\label{eq:edge_forwarding}
    \mathbf{e}_{vu}^{(l+1)} = g_\Theta^{(l)}\left(\mathbf{e}_{vu}^{(l)}\right)
\end{equation}



\subsection{Node Operation}
To complete the explanation of our convolutional block, we specify the specific computation for the generic node operation $\psi$ introduced in Equation~\ref{eq:psi}:

\begin{equation}
    \bm{h}_{u}^{(l+1)} = \bm{W}_1^{(l)} \textrm{cat}\left( \bm{h}^{(l)}_u, 
\sum_{v \in \mathcal{N}(u)} \bm{\eta}^{(l)}_{vu} \right)\label{eq:our_mespas}\\
\end{equation}

\noindent
The aggregation function is a sum, and we choose to concatenate the hidden representation of $u$ with the messages $\eta_{vu}$ from the other nodes.
For heterogeneous convolutions (i.e. Equation~\ref{eq:aggregation_heterogeneous}), we use $\textrm{mean}(\cdot)$ as the aggregation function.  
Such choices are not motivated by musical reasoning, and experimenting with other operations could be an interesting direction, but are out of the scope of this paper.

\section{Data}\label{sec:datasets}



Musical score graph datasets are different from common datasets in the graph-related literature regarding the number and size of graphs.
Node classification and link prediction datasets often only consist of a single huge graph, coupled with a sampling strategy to obtain subgraphs to train and evaluate the Graph Neural Network~\cite{hamilton2017inductive,zou2019layer}. On the other hand, graph classification datasets often consist of a large number of small graphs, with less than 50 nodes~\cite{gnnsurvey_tnnls21}. 
Musical score graphs are neither small nor extremely large and may vary significantly in size; a Bach Chorale may have $\sim100$ notes whereas a Beethoven Sonata might have more than $5000$ (with every note corresponding to a node in the graph). 

Moreover, popular sampling strategies, such as node-wise sampling, subgraph sampling, random-walk sampling, and spectral sampling, may yield musically problematic note configurations, for example, by segregating notes that are played at the same time while grouping in the same subgraph notes that are very far apart. This is problematic, especially for musically-local tasks such as voice separation or Roman numeral analysis, where the system cannot be expected to produce a meaningful result, and could even learn to perform the task in the wrong way if some notes (and onsets for the roman numeral analysis) are missing.

In the previous works on graph scores that we are considering, the problem is avoided by training on mini-batches which consist of single pieces. However, this is not an efficient solution since it always leaves a big part of memory unused, thus unnecessarily prolonging the training time and decreasing the variability in the batch. We describe the new sampling mechanism we use in the following section, and then move on to detailing the datasets used in our experiments.

\subsection{Data Sampling}
While our nodes can be ordered by multiple features, the organizational aspect that is most prominent from a perceptual point of view is time. Indeed, people would still recognize a music piece if it is segmented over the time axis, while, for example, considering only pitches in a certain interval could lead to meaningless results. There is also perceptual evidence that the offset time of a note is much less salient than the onset time, especially for percussive instruments (including the piano) whose sound naturally decays over time~\cite{klapuri}.
Therefore, when we create our graphs from a musical score, we set the node order first by the absolute time of onset and then by pitch.

Once this ordering is set (and having defined an index function $ind(\cdot)$ that given a node returns its index in this ordering)
we randomly sample a subgraph of size $K > 1$ from a piece with $N$ notes, with the following procedure. If $N>K$ we select the nodes $u$ with $ind(v) \leq ind(u) \leq ind(v)+N$, where $v$ is a random node which satisfies the inequality $ind(v)<K-N$. If $N \leq K$ we select all nodes in the piece. 
Note that we can still have the problem of segregating notes from the same chord, but this can now only happen at the temporal boundaries of our subgraph, limiting its impact on the network.

We can then create a batch consisting of $B$ subgraphs of at most size $N$. Our batching mechanism uses the approach of Hamilton et al.~\shortcite{hamilton2017inductive},
i.e., all graphs are joined together in a single batched graph that will contain $B$ disjoint subgraphs.


\subsection{Datasets}
We use four distinct datasets for our four tasks.

\subsubsection{Voice Separation}
The Graph Voice Separation dataset was introduced by Karystinaios et al.~\shortcite{karystinaios2023voice}. This dataset contains graph data created from five collections: 370 Bach Chorales, 48 Preludes and 48 Fugues from the Bach Well-Tempered Clavier (Books I and II), 15 Bach Inventions, 15 Bach Sinfonias, and 210 movements from Haydn String Quartets. It contains in total $726,246$ nodes and $3,408,679$ edges from $1,054$ unique score graphs. Karystinaios et al. only test on single collections to understand the differences in performance for different composing styles. To have a single general performance indicator, we introduce a new data split that uses $70$\% of the data for training, $10$\% for validation, and $20$\% for testing. This split preserves the percentage of pieces in each collection and is independent of the size of each score graph.

\subsubsection{Composer Clasification}
For composer classification, we use the scores from the DCML corpora dataset\footnote{\url{https://github.com/DCMLab/dcml_corpora}}. 
The dataset includes 10 composers, for a total of 419 scores, from where we build score graphs with collectively $710,240$ nodes and $3,924,655$ edges.  We create a random data split with $70$\% of the data for training, $10$\% for validation, and $20$\% for testing, which preserves the percentage of composers in each set.

\subsubsection{Roman Numeral Analysis}
The Roman numeral analysis dataset with data augmentation was introduced by Lopez et al.~\shortcite{lopez2021augmentednet}. We use their dataset (with augmentations on the train set). The created graphs collectively contain $8,968,413$ nodes, $38,390,729$ edges, and $5,096,853$ unique onset positions from $7,988$ scores (after data augmentation).

\subsubsection{Cadence Detection}
For the Cadence Detection task, we use four distinct annotated datasets, the Mozart Piano Sonatas~\cite{Hentschel2021MozartPianoSonatas}, Haydn String Quartets~\cite{sears2017haydn_string_quartets}, Mozart String Quartets~\cite{allegraud2019mozart_string_quartets}, and Bach WTC Fugues~\cite{giraud2015bachWTC}. The created graph collectively contains $300,602$ nodes and $1,392,753$ edges from $153$ scores. We create a random data split with $70$\% of the data for training, $10\%$ for validation, and $20$\% for testing.

\section{Experiments}\label{sec:experiments}
Our model for each of the four tasks is built on the respective current state-of-the-art model presented in Section~\ref{sec:tasks}. 
From here on forward, we refer to the previous state-of-the-art architecture as \textit{baseline} model, and to the same architecture with the convolutional blocks replaced with our new ones, as \textit{\our model}. These original models serve as the baselines in the following experiment. We follow the implementations of the publicly available code with no major modifications, except for the sampling technique that we highlighted before.
For each task, we consider the main evaluation metric proposed in the original paper (presented in Section~\ref{sec:tasks}).

All experiments for a certain task are run with the best hyperparameter setting specified in the respective papers; this includes 2 GNN layers, and the convolutional blocks being SageConv~\cite{hamilton2017inductive} for all tasks except the voice separation where ResGatedGraphConv~\cite{Bresson2017ResidualGG} is used.
We use a fixed training, validation, and test split for each task, and every experiment is run 10 times with different NN initialisations on a single GPU. We used one GTX 1080 Ti GPU with 11 GB of VRAM.



\subsection{Main Results}\label{subsec:results}
    


\begin{table*}[]
    \centering
    \begin{tabular}{r|c c c c}
        \toprule
         &  Voice Separation & Composer clf & RNA & Cadence Detection\\
         &  (Link Prediction) & (Graph Classification) & (Node Classification) & (Node Classification)\\
         \midrule
         Previous SOTA Arch & $0.8111\pm0.058$& $0.4288\pm0.031$& $0.3221\pm0.010$& $0.4065\pm0.011$\\
         \hline
         MusGConv                   & $0.8142\pm0.035$& $\mathbf{0.5233}\pm0.032$& $0.3126\pm0.015$& $0.4126\pm0.016$\\
         MusGConv(+EF)              & $\mathbf{0.8436}\pm0.032$& $0.3939\pm0.018$& $0.3177\pm0.010$& $\mathbf{0.4295}\pm0.009$\\
         \bottomrule
    \end{tabular}
    \caption{Experimental comparison with  previous SOTA models.
    The evaluation metric varies for the different tasks (see the corresponding subsections in Section \ref{sec:tasks}). Marked in bold are the best results when they are statistically significant.}
    \label{tab:main_exp}
\end{table*}

The goal of our main experiment is to quantitatively verify whether the use of \our can improve the results on the four tasks compared to the respective baseline, i.e., the architectures used in the corresponding state-of-the-art approaches. 
Each experiment is run 10 times with the fixed task-specific data split (as described in the previous section) and different random seeds.
We report ASO significance~\cite{del2018optimal} with a confidence level $\alpha = 0.05$ and $\epsilon_\text{min} < 0.1$. 

The results, summarised in Table~\ref{tab:main_exp}, show that \ouref produces statistically significant better results for the Voice Separation and Cadence Detection task. In the Composer Classification task, the best-performing model is \our, while \ouref yields worse results than the baseline. We suspect that being this a global graph classification task, the edge features propagation is harder to train, and the simpler mechanism employed in \our is a better choice. There is no statistically significant difference in the performance on the Roman Numeral Analysis task. We found two potential explanations for this behaviour: first, the RNA model is very complex, with multiple components which could hide the effect of a modification on the graph encoder. Moreover, the RNA dataset is augmented with transpositions in all keys, and therefore having transposition-invariant features may only yield minimal (if any) advantage. An experiment without augmentation is not possible, since the output of the RNA model depends on the absolute pitches at the input, and the not-augmented dataset is very small.

\begin{figure}[tb]
\centerline{\includegraphics[width=0.85\columnwidth]{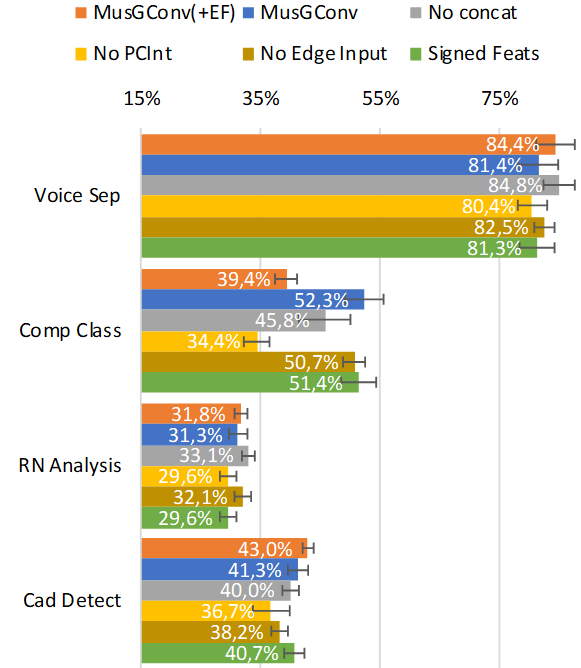}}
\caption{Ablation studies.}
\label{fig:ablation}
\end{figure}

Regarding execution time, for each task, we compute the ratio between the average time of the 10 runs for baseline and the 10 runs with \our and \ouref. Aggregated across all tasks, this ratio has an average of $1 \pm 0.05$. Thus, the usage of \our has a minimal impact on the final execution time.


\subsection{Ablation Studies}\label{subsec:ablation}
We conduct four ablation studies to explore different model variations in terms of architecture and selection of edge features. The results are reported in Figure~\ref{fig:ablation}, where we also include the results for \our and \ouref from our main experiment (see previous section) for comparison. For some variations we consider, there is not a clear winner, meaning that, while the usage of relative features as edges is beneficial overall, different versions of our system can perform better on different musical tasks.

\paragraph{No concatenation.} We change the feature aggregation function $\phi$ (Eq.~\ref{eq:eta}) from a concatenation (Eq.~\ref{eq:our_eta}) to a multiplication, to mimic the operation used in convolutional blocks that deal with edge features, e.g., \cite{Atzmon2018,WangSLSBS19,Satorras2021EnEG}). 
The results show that this degrades the performance for the composer classification and cadence detection tasks while improving RN analysis and Voice Separation. 

\paragraph{No Edge Input.}
In this study, we ignore our manually built edge features from Section~\ref{sec:edge_feat_comp} and use node feature difference (see Eq.~\ref{eq:edgesub}) as edge features for all blocks (including the first). This is similar to the edge features employed in EdgeConv~\cite{wang2019dynamic} (though it has to be noted that our message passing is different from EdgeConv). This degrades the model performance on all tasks but RN.

\paragraph{No PCInt.}
We quantify the effect of the PCInt edge feature as a much more music-specific edge feature, compared to the ``more standard'' feature distances.
We observe that the usage of this feature improves all tasks.


\paragraph{Signed Features.}
We evaluate the use of features obtained by removing the absolute value operation in Eq.~\ref{eq:relational_features}. Indeed this is a more informative input; for example, with the absolute value we encode the difference between two note durations, but we lose the information about which is longer. On the other hand, it increases the input numerical range, and one could argue that the network already has access to absolute features, and PCInt and edge type encode similar information.
The results show that using signed features is not beneficial. 

\section{Conclusion and Future Work}\label{sec:conclusion}

This paper has presented a graph convolution block dedicated to music understanding tasks. Its working mechanism is inspired by perceptual considerations and permits the propagation of transposition-invariant and relative timing information in the message-passing process. More generally, our work enables an elegant and efficient way to use pairwise note features, which have been long studied and employed in monophonic music, for polyphonic music processing. Specifically, our approach can be summarized in two core contributions: pitch and time pairwise functions as edge features, and a new way of aggregating this information inside the convolutional block. The design of this block is kept simple to give us a performance advantage without increasing computation time. We experimentally verify the validity of our proposition on four rather diverse musical tasks covering three graph-related problems: graph classification, node classification, and link prediction. 

As future work, it would be interesting to evaluate \our on other kind of music and to investigate the impact of other pairwise note functions, like the one proposed in the cognitively plausible music model IDyOM~\cite{Pearce:2018,Pearce:idyom}.

\section*{Acknowledgements}
This work was supported by the European Research Council (ERC) under the EU's Horizon 2020 research \& innovation programme, grant agreement No.\ 101019375 (\textit{Whither Music?}), and the Federal State of Upper Austria (LIT AI Lab).

    

\bibliographystyle{named}
\bibliography{ijcai24}

\begin{thebibliography}{}

\bibitem[\protect\citeauthoryear{Allegraud \bgroup \em et al.\egroup }{2019}]{allegraud2019mozart_string_quartets}
Pierre Allegraud, Louis Bigo, Laurent Feisthauer, Mathieu Giraud, Richard Groult, Emmanuel Leguy, and Florence Lev{\'e}.
\newblock Learning sonata form structure on mozart's string quartets.
\newblock {\em {Transactions of the International Society for Music Information Retrieval (TISMIR)}}, 2(1):82--96, 2019.

\bibitem[\protect\citeauthoryear{Atzmon \bgroup \em et al.\egroup }{2018}]{Atzmon2018}
Matan Atzmon, Haggai Maron, and Yaron Lipman.
\newblock Point convolutional neural networks by extension operators.
\newblock {\em {ACM} Transactions on Graphics}, 37(4), 2018.

\bibitem[\protect\citeauthoryear{Bresson and Laurent}{2017}]{Bresson2017ResidualGG}
Xavier Bresson and Thomas Laurent.
\newblock Residual gated graph convnets.
\newblock In {\em Proceedings of the International Conference on Learning Representations}, 2017.

\bibitem[\protect\citeauthoryear{Chawla \bgroup \em et al.\egroup }{2002}]{chawla2002smote}
Nitesh~V Chawla, Kevin~W Bowyer, Lawrence~O Hall, and W~Philip Kegelmeyer.
\newblock Smote: synthetic minority over-sampling technique.
\newblock {\em Journal of artificial intelligence research}, 16:321--357, 2002.

\bibitem[\protect\citeauthoryear{Chen \bgroup \em et al.\egroup }{2018}]{chen2018functional}
Tsung-Ping Chen, Li~Su, et~al.
\newblock {Functional Harmony Recognition of Symbolic Music Data with Multi-task Recurrent Neural Networks.}
\newblock In {\em {Proceedings of the International Society for Music Information Retrieval Conference (ISMIR)}}, 2018.

\bibitem[\protect\citeauthoryear{Del~Barrio \bgroup \em et al.\egroup }{2018}]{del2018optimal}
Eustasio Del~Barrio, Juan~A Cuesta-Albertos, and Carlos Matr{\'a}n.
\newblock An optimal transportation approach for assessing almost stochastic order.
\newblock In {\em The Mathematics of the Uncertain}, pages 33--44. Springer, 2018.

\bibitem[\protect\citeauthoryear{Deutsch}{2013}]{deutsch2013psychology}
Diana Deutsch.
\newblock {\em Psychology of music}.
\newblock Elsevier, 2013.

\bibitem[\protect\citeauthoryear{Di~Giorgi \bgroup \em et al.\egroup }{2021}]{Giorgi2021DownbeatTW}
Bruno Di~Giorgi, Matthias Mauch, and Mark Levy.
\newblock Downbeat tracking with tempo-invariant convolutional neural networks.
\newblock In {\em Proceedings of the International Society for Music Information Retrieval Conference (ISMIR)}, 2021.

\bibitem[\protect\citeauthoryear{Duane and Pardo}{2009}]{duane2009streaming}
Ben Duane and Bryan Pardo.
\newblock Streaming from midi using constraint satisfaction optimization and sequence alignment.
\newblock In {\em Proceedings of the International Computer Music Conference (ICMC)}, 2009.

\bibitem[\protect\citeauthoryear{Elowsson and Friberg}{2019}]{Elowsson2019ModelingMM}
Anders Elowsson and Anders Friberg.
\newblock {Modeling Music Modality with a Key-Class Invariant Pitch Chroma CNN}.
\newblock In {\em {Proceedings of the International Society for Music Information Retrieval Conference (ISMIR)}}, 2019.

\bibitem[\protect\citeauthoryear{Foscarin \bgroup \em et al.\egroup }{2021}]{foscarin2021pkspell}
Francesco Foscarin, Nicolas Audebert, and Rapha{\"e}l Fournier-S'Niehotta.
\newblock Pkspell: Data-driven pitch spelling and key signature estimation.
\newblock In {\em {Proceedings of the International Society for Music Information Retrieval Conference (ISMIR)}}, 2021.

\bibitem[\protect\citeauthoryear{Foscarin \bgroup \em et al.\egroup }{2022}]{foscarin2022match}
Francesco Foscarin, Emmanouil Karystinaios, Silvan~David Peter, Carlos Cancino-Chac{\'o}n, Maarten Grachten, and Gerhard Widmer.
\newblock The match file format: Encoding alignments between scores and performances.
\newblock In {\em Proceedings of the Music Encoding Conference ({MEC})}, 2022.

\bibitem[\protect\citeauthoryear{Fradet \bgroup \em et al.\egroup }{2021}]{miditok2021}
Nathan Fradet, Jean-Pierre Briot, Fabien Chhel, Amal El~Fallah~Seghrouchni, and Nicolas Gutowski.
\newblock {MidiTok}: A python package for {MIDI} file tokenization.
\newblock In {\em Late-Breaking Demo Session of the International Society for Music Information Retrieval Conference (ISMIR)}, 2021.

\bibitem[\protect\citeauthoryear{Giraud \bgroup \em et al.\egroup }{2015}]{giraud2015bachWTC}
Mathieu Giraud, Richard Groult, Emmanuel Leguy, and Florence Lev{\'e}.
\newblock {Computational Fugue Analysis}.
\newblock {\em {Computer Music Journal}}, 39(2):77--96, 2015.

\bibitem[\protect\citeauthoryear{Hamilton \bgroup \em et al.\egroup }{2017}]{hamilton2017inductive}
Will Hamilton, Zhitao Ying, and Jure Leskovec.
\newblock Inductive representation learning on large graphs.
\newblock {\em Advances in Neural Information Processing Systems}, 30, 2017.

\bibitem[\protect\citeauthoryear{Harte}{2010}]{harte2010towards}
Christopher Harte.
\newblock {\em Towards automatic extraction of harmony information from music signals}.
\newblock PhD thesis, Queen Mary, University of London, 2010.

\bibitem[\protect\citeauthoryear{Hentschel \bgroup \em et al.\egroup }{2021}]{Hentschel2021MozartPianoSonatas}
Johannes Hentschel, Markus Neuwirth, and Martin Rohrmeier.
\newblock The annotated mozart sonatas: Score, harmony, and cadence.
\newblock {\em {Transactions of the International Society for Music Information Retrieval (TISMIR)}}, 4(1), May 2021.

\bibitem[\protect\citeauthoryear{Hsiao and Su}{2021}]{hsiao2021learning}
Yo-Wei Hsiao and Li~Su.
\newblock Learning note-to-note affinity for voice segregation and melody line identification of symbolic music data.
\newblock In {\em {Proceedings of the International Society for Music Information Retrieval Conference (ISMIR)}}, pages 285--292, 2021.

\bibitem[\protect\citeauthoryear{Hu \bgroup \em et al.\egroup }{2019}]{hu2019strategies}
Weihua Hu, Bowen Liu, Joseph Gomes, Marinka Zitnik, Percy Liang, Vijay Pande, and Jure Leskovec.
\newblock Strategies for pre-training graph neural networks.
\newblock In {\em International Conference on Learning Representations (ICLR)}, 2019.

\bibitem[\protect\citeauthoryear{Jeong \bgroup \em et al.\egroup }{2019}]{jeong2019graph}
Dasaem Jeong, Taegyun Kwon, Yoojin Kim, and Juhan Nam.
\newblock Graph neural network for music score data and modeling expressive piano performance.
\newblock In {\em {Proceedings of the International Conference on Machine Learning (ICML)}}, pages 3060--3070. PMLR, 2019.

\bibitem[\protect\citeauthoryear{Karystinaios and Widmer}{2022}]{karystinaios2022cadence}
Emmanouil Karystinaios and Gerhard Widmer.
\newblock Cadence detection in symbolic classical music using graph neural networks.
\newblock In {\em Proceedings of the International Society for Music Information Retrieval Conference (ISMIR)}, 2022.

\bibitem[\protect\citeauthoryear{Karystinaios and Widmer}{2023}]{karystinaios2023roman}
Emmanouil Karystinaios and Gerhard Widmer.
\newblock {Roman Numeral Analysis with Graph Neural Networks: Onset-wise Predictions from Note-wise Features}.
\newblock In {\em {Proceedings of the International Society for Music Information Retrieval Conference (ISMIR)}}, 2023.

\bibitem[\protect\citeauthoryear{Karystinaios \bgroup \em et al.\egroup }{2023}]{karystinaios2023voice}
Emmanouil Karystinaios, Francesco Foscarin, and Gerhard Widmer.
\newblock {Musical Voice Separation as Link Prediction: Modeling a Musical Perception Task as a Multi-Trajectory Tracking Problem}.
\newblock In {\em {Proceedings of the Joint Conference on Atrificial Intelligence (IJCAI)}}, 2023.

\bibitem[\protect\citeauthoryear{Kermarec \bgroup \em et al.\egroup }{2022}]{kermarec2022improving}
Mathieu Kermarec, Louis Bigo, and Mikaela Keller.
\newblock Improving tokenization expressiveness with pitch intervals.
\newblock In {\em Late-Breaking Demo Session of the International Society for Music Information Retrieval Conference (ISMIR)}, 2022.

\bibitem[\protect\citeauthoryear{Klapuri and Davy}{2006}]{klapuri}
Anssi Klapuri and Manuel Davy.
\newblock {\em Signal Processing Methods for Music Transcription}.
\newblock Springer, 2006.

\bibitem[\protect\citeauthoryear{Lattner \bgroup \em et al.\egroup }{2018}]{Lattner2018LearningTI}
Stefan Lattner, Maarten Grachten, and Gerhard Widmer.
\newblock {Learning Transposition-Invariant Interval Features from Symbolic Music and Audio}.
\newblock In {\em {Proceedings of the International Society for Music Information Retrieval Conference (ISMIR)}}, 2018.

\bibitem[\protect\citeauthoryear{Lattner \bgroup \em et al.\egroup }{2019}]{Lattner2019LearningCB}
Stefan Lattner, Monika D{\"o}rfler, and Andreas Arzt.
\newblock {Learning Complex Basis Functions for Invariant Representations of Audio}.
\newblock In {\em {Proceedings of the International Society for Music Information Retrieval Conference (ISMIR)}}, 2019.

\bibitem[\protect\citeauthoryear{L{\'o}pez \bgroup \em et al.\egroup }{2021}]{lopez2021augmentednet}
N{\'e}stor~N{\'a}poles L{\'o}pez, Mark Gotham, and Ichiro Fujinaga.
\newblock {AugmentedNet: A Roman Numeral Analysis Network with Synthetic Training Examples and Additional Tonal Tasks.}
\newblock In {\em {Proceedings of the International Society for Music Information Retrieval Conference (ISMIR)}}, 2021.

\bibitem[\protect\citeauthoryear{McLeod and Rohrmeier}{2021}]{mcleod2021modular}
Andrew~Philip McLeod and Martin~Alois Rohrmeier.
\newblock {A modular system for the harmonic analysis of musical scores using a large vocabulary}.
\newblock In {\em {Proceedings of the International Society for Music Information Retrieval Conference (ISMIR)}}, 2021.

\bibitem[\protect\citeauthoryear{Micchi \bgroup \em et al.\egroup }{2020}]{micchi2020not}
Gianluca Micchi, Mark Gotham, and Mathieu Giraud.
\newblock {Not all roads lead to Rome: Pitch representation and model architecture for automatic harmonic analysis}.
\newblock {\em {Transactions of the International Society for Music Information Retrieval (TISMIR)}}, 3(1):42--54, 2020.

\bibitem[\protect\citeauthoryear{Nakamura \bgroup \em et al.\egroup }{2016}]{nakamura2016tree}
Eita Nakamura, Masatoshi Hamanaka, Keiji Hirata, and Kazuyoshi Yoshii.
\newblock Tree-structured probabilistic model of monophonic written music based on the generative theory of tonal music.
\newblock In {\em Proceedings of the International Conference on Acoustics, Speech and Signal Processing (ICASSP)}, pages 276--280. IEEE, 2016.

\bibitem[\protect\citeauthoryear{Pearce}{2005}]{Pearce:idyom}
Marcus Pearce.
\newblock {\em {The Construction and Evaluation of Statistical Models of Melodic Structure in Music Perception and Composition}}.
\newblock PhD thesis, City University of London, UK, 2005.

\bibitem[\protect\citeauthoryear{Pearce}{2018}]{Pearce:2018}
Marcus~T. Pearce.
\newblock Statistical learning and probabilistic prediction in music cognition: Mechanisms of stylistic enculturation.
\newblock {\em Annals of the New York Academy of Sciences}, 1423(1):378--395, 2018.

\bibitem[\protect\citeauthoryear{Satorras \bgroup \em et al.\egroup }{2021}]{Satorras2021EnEG}
Victor~Garcia Satorras, Emiel Hoogeboom, and Max Welling.
\newblock E(n) equivariant graph neural networks.
\newblock In {\em {Proceedings of the International Conference on Machine Learning (ICML)}}, 2021.

\bibitem[\protect\citeauthoryear{Schlichtkrull \bgroup \em et al.\egroup }{2018}]{SchlichtkrullKB18}
Michael~Sejr Schlichtkrull, Thomas~N. Kipf, Peter Bloem, Rianne van~den Berg, Ivan Titov, and Max Welling.
\newblock Modeling relational data with graph convolutional networks.
\newblock In {\em Proceedings of the Semantic Web International Conference, {ESWC}}, volume 10843 of {\em Lecture Notes in Computer Science}, pages 593--607. Springer, 2018.

\bibitem[\protect\citeauthoryear{Sears \bgroup \em et al.\egroup }{2017}]{sears2017haydn_string_quartets}
David~RW Sears, Andreas Arzt, Harald Frostel, Reinhard Sonnleitner, and Gerhard Widmer.
\newblock Modeling harmony with skip-grams.
\newblock In {\em {Proceedings of the International Society for Music Information Retrieval Conference (ISMIR)}}, 2017.

\bibitem[\protect\citeauthoryear{Wang \bgroup \em et al.\egroup }{2019a}]{WangSLSBS19}
Yue Wang, Yongbin Sun, Ziwei Liu, Sanjay~E. Sarma, Michael~M. Bronstein, and Justin~M. Solomon.
\newblock Dynamic graph {CNN} for learning on point clouds.
\newblock {\em {ACM} Transactions on Graphics}, 38(5):146:1--146:12, 2019.

\bibitem[\protect\citeauthoryear{Wang \bgroup \em et al.\egroup }{2019b}]{wang2019dynamic}
Yue Wang, Yongbin Sun, Ziwei Liu, Sanjay~E Sarma, Michael~M Bronstein, and Justin~M Solomon.
\newblock Dynamic graph cnn for learning on point clouds.
\newblock {\em ACM Transactions on Graphics (tog)}, 38(5):1--12, 2019.

\bibitem[\protect\citeauthoryear{Wu \bgroup \em et al.\egroup }{2021}]{gnnsurvey_tnnls21}
Zonghan Wu, Shirui Pan, Fengwen Chen, Guodong Long, Chengqi Zhang, and Philip~S. Yu.
\newblock A comprehensive survey on graph neural networks.
\newblock {\em IEEE Transactions on Neural Networks and Learning Systems}, 32(1):4--24, 2021.

\bibitem[\protect\citeauthoryear{Zhang \bgroup \em et al.\egroup }{2023}]{huan}
Huan Zhang, Emmanouil Karystinaios, Carlos Cancino-Chac{\'o}n, Simon Dixon, and Gerhard Widmer.
\newblock Symbolic music representations for classification tasks: A systematic evaluation.
\newblock In {\em {Proceedings of the International Society for Music Information Retrieval Conference (ISMIR)}}, 2023.

\bibitem[\protect\citeauthoryear{Zou \bgroup \em et al.\egroup }{2019}]{zou2019layer}
Difan Zou, Ziniu Hu, Yewen Wang, Song Jiang, Yizhou Sun, and Quanquan Gu.
\newblock Layer-dependent importance sampling for training deep and large graph convolutional networks.
\newblock {\em Advances in neural information processing systems}, 32, 2019.

\end{thebibliography}

\end{document}